# Do Two Symmetry-Breaking Transitions in Photosynthetic Light Harvesting Complexes (PLHC) Form One, Two or More Kibble-Zurek Model (KZM) Topological Defect(s)?

N. H. March[1], R. H. Squire[2]

[1] Department of Physics, University of Antwerp, Belgium
Oxford University, Oxford, England
[2] Department of Natural Sciences, West Virginia University - Institute of Technology, Beckley, WV, 25136, USA

**Abstract**

Kibble [1] and Zurek [2] proposed that rapid symmetry-breaking transitions in the hot, early universe could result in causally disconnected topological defects such as cosmic strings. This type of first order transition has analogues in certain second order transitions present in condensed matter such as liquid crystals, superfluids and charge density waves in terms of flux tubes or vortices. Recent, we discovered that Rhodopseudomonas acidophila's Photosynthetic Light Harvesting Complex might have different types of coherent ground and excited states, suggesting that there are two different symmetry-breaking transitions. The B850 ground states comprise eight identical rings each containing 18 bacteriochlorophyll components, and each ring has undergone a Bose-Einstein phase transition to a charge density wave that lowers the energy. The excited state coherence results from polariton formation from the non-crossing of bosons, here excitons and photons, an extension of exciton theory [3]. The result is short-lived quasi-particles with very low mass that can form an unusual BEC. We suggest the oriented, circular B850's and enclosed single B875 create new cavity structure with some attributes of toroidal nanopillars [4, 5]. Since both the ground and excited states should contain solitons, we envisage three fast light pulse experiments should be able to map both the KZM phase transitions and energy transfers as a function of light intensity and time in this complex at room temperature [6, 7].

**1. Introduction.** Several years ago, Professor March and I independently studied the photosynthetic process and then we concluded that the efficiency of energy transfer was beyond our understanding. What we are offering here is a model that rationalizes how coherent energy transfer from light absorption to the reaction center (RC) might proceeds. Though the PLHC ring has been studied for many years, it apparently has not been described as a CDW, despite the "dimer" feature (confirmed by crystallography) as described in section 2. Section 3a discusses symmetry breaking, 3b outlines the KZM theory, and 3c outlines coherence, followed by 3d, the Geodesic Rule (phase addition) and defect types. Section 5 is devoted to polaritons. Certainly, recent advances in optical micro- and nanocavity spectroscopy have permitted a deluge of significant experiments identifying the properties of excited states of CDW's and polaritons, despite their transient nature. Finally, we put the pieces together (section 6) in a coherent fashion and discuss the cooperation that may be present to transfer energy efficiently, followed by a summary, conclusions, and possibilities for further work (section 7).

## 2. Brief Description of Photosynthetic Light Harvesting Complexes (PLHC) [8, 9, 10].

In anaerobic photosynthetic prokaryotes (purple bacteria), the process of light capture and transfer is highly efficient. Our emphasis is on capture of solar energy and its nearly perfect transfer to the reaction center (RC or P870). A purple bacterium (Rhodopseudomonas acidophila) has a photosynthetic unit comprised of two pigment-protein light-harvesting (LH) complexes called LH1/RC and LH2. These LH's have similar molecular structures comprised of bacteriochlorophyll components (BChl a, designated BChl hereafter) that are comprised of circular conjugated arrays with a coordinated magnesium ion, Mg (2+), Fig 1.

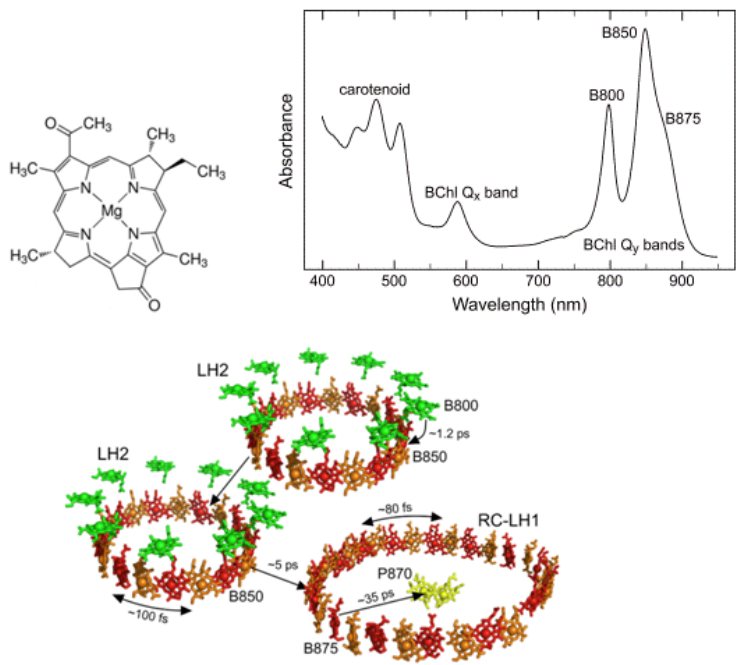

Figure 1a (top left) A prototypical bacteriochlorophyll a (BChl)$_a$; (right) Spectrum of BChl complexes B800 (green below), B850 and B875 (red and yellow). The Soret bands are located between 300 – 400 nm. 1b. Graphical illustration of crystal structure coordinates for light harvesting (LH) complexes 1, 2 and the reaction center (P870). The orange, red and green structures are comprised of the BChl ring structure surrounding a Mg(+2) that is responsible for the charge density wave that generates the "dimer" structure of the B850 and B875 rings (see text). The absorption maxima indicate the complexes; EET rates are shown. Figure courtesy of Professor M. Jones (Bristol).

Eight LH2's, each with 18 BChl units surround the RC-LH1. X-ray crystallography has determined the atomic structure and orientations [6,7], and the associated pigments groups are separated clearly enough so electronic energy transfer (EET) can be measured. The complexes' objective is to absorb and transfer energy efficiently to the reaction center (RC) where "charge splitting" takes place to produce a 1.1eV potential and thus generate energy for the survival and growth of the bacteria. The oxygen-generating bacteria are presumed to be responsible for converting the oceans and atmosphere from the $CO_2$ environment to one rich in oxygen (the "Great Oxygenation Event") a couple of billion years ago [11].

**3. Background. a. Spontaneous Symmetry Breaking (SSB).** SSB is ubiquitous in our daily life; water freezes, and the rotational symmetry is broken by a selection of one out of many rotational possibilities by the newly formed ice crystal. Many crystals in a proper environment can undergo a very slow crystallization to minimize defects. However, in the early universe, the transition was rapid and multiple stable topological defects should have been formed. The stages of the development of the "wine bottle" SSB potential (fig 3b) are formed by sufficiently lowering the temperature (fig 3a).

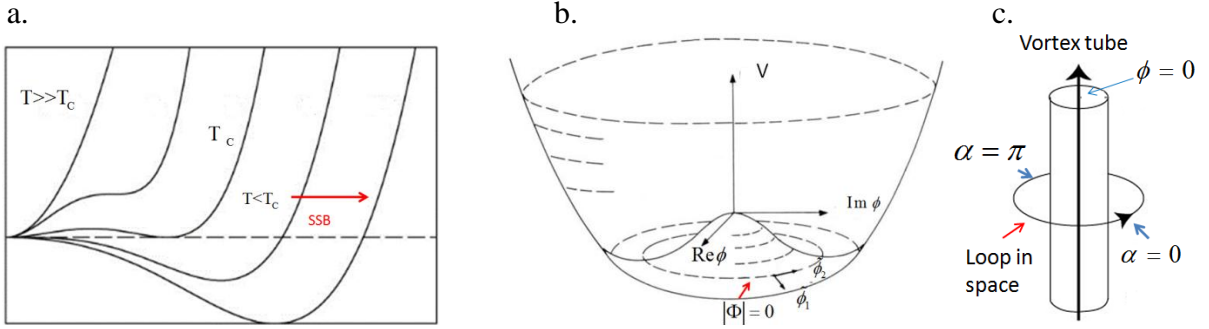

Figure 2. a. Rapid cooling of the universe changes the parabola potential to the "wine bottle" structure by rotating one of the two potentials (rhs of a) around V (3b). b. The choice of phase is arbitrary, but once chosen, this state can generate a Bose-Einstein condensation if certain parameters are suitable (temperature, mass, etc.; see coherence section later). c. The simplest order parameter is complex, $\Phi$. If we have a system with three separated regions, each with an arbitrary phase, and we trace a large loop in space and add the phase values, if the phase changes by $2n\pi$, there must be at least one vortex present (2c.). The number n is then "topological".

The potential $V(\psi) = \lambda\left(|\phi|^2 - \rho^2\right)^2$ is symmetric about the rotation axis, i.e. $\psi \to \psi e^{i\phi}$ about $V(\phi=0)$, $\lambda$ and $\rho$ are constants. At $T \gg T_c$, thermal fluctuations are much more energetic that the $V(\phi)$ and the average $\phi = 0$. When $T < T_c$, an arbitrary phase is selected and the symmetry is broken and the minima occur on the circle $|\Phi|$ with the same energy because of symmetry, but are different ground states since their phase angles differ in general. However, a sizable fraction of, say, cold atoms with the same phase angle can create a Bose condensation if the temperature and/or mass is low enough. Both CDW's and polaritons break the electromagnetic field symmetry, so once $\phi$ has been selected, SSB prevails: $U(1) = e^{i\phi}; (\phi = 0 \to 2\pi)$.

**3b. Coherence.** A Hamiltonian describing bosons such as excitons is

$$H = \sum_i \frac{\hbar^2 k^2}{2m} \hat{a}_k^\dagger \hat{a}_k$$

where $\hat{a}_k^\dagger, \hat{a}_k$ are the creation and annihilation operators, $k$ is the wave vector, and $m$ is the mass. The distribution number of *bosonic* particles is different from fermions (-1 is changed to +1 for fermions) with $\mu$, the chemical potential and $\beta = 1/k_B T$, where $k_B$ is the Boltzmann constant,

$$N_{\vec{k}} = \frac{1}{e^{\beta(\hbar^2 k^2/2m - \mu)} - 1} \qquad (1)$$

In the thermodynamic limit (large number of particles), the density of particles, n, is

$$n = \lim_{V \to \infty} \frac{N}{V} = \frac{1}{V}\sum_{\vec{k}} N_{\vec{k}} = \frac{1}{(2\pi)^3}\int d^3k \frac{1}{e^{\beta(\hbar^2 k^2/2m)}-1} \quad (2)$$

so, the integral becomes [12, 13]

$$n = 2.612\frac{(mk_B T)^{3/2}}{(2\pi\hbar^2)^{3/2}} \quad \text{or} \quad T_c = \frac{2\pi\hbar^2}{k_B m}\left(\frac{n}{2.612}\right)^{2/3} \quad (3)$$

However, the integral has not accounted for all the particles since the zero state, $n_0$, must be treated separately, so rewriting the density eq. (2)

$$n = \lim_{V \to \infty}\frac{N}{V} = n_0 + \frac{1}{(2\pi)^3}\int d^3k \frac{1}{e^{\beta(\hbar^2 k^2/2m)}-1} \quad (2')$$

When a large number of bosons are cooled below $T_c$ (*note the inverse dependence on mass* (eq. (3), a large fraction of the bosons occupy the lowest state and become coherent which means that the phase, $\varphi$, is no longer random ($0 \to 2\pi$), but becomes a single value and each boson "phase locks" with all of the other bosons as a result of interactions. Identifying bosons as excitons would mean that we could diagonalize the exciton density matrix in a new basis and represent all the participating excitons with a single wave function, $\psi(\vec{r},t) = \sqrt{n_0(\vec{r},t)}e^{iS(\vec{r},t)}$, where $n_0$ was defined by eq. (2') and $S(r,t)$ is the common phase. The "normal" state phases of a wave function are arbitrary from $0 \to 2\pi$. In a BEC this phase value is arbitrary, but once chosen, the electromagnetic field symmetry has been "broken". Further, the chemical potential has now become zero, so the "resistance" to adding or removing a particle is now zero (see fig. 3).

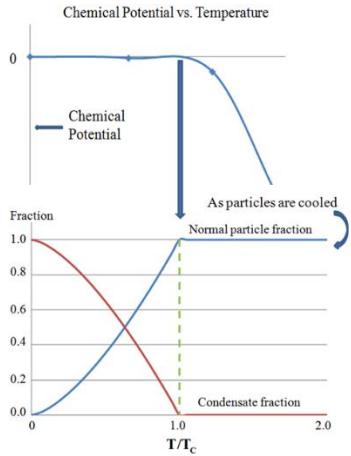

Figure 3. When the chemical potential reaches zero, $T_c$ is defined and Bose condensation begins. Thus, the energy cost to add or remove a particle in a BEC is zero, and the uncertainty principle is written as $\Delta N \Delta \varphi \sim \hbar$, where $\Delta \varphi$ is a single value, so $\Delta N$ has some flexibility.

The physical significance of the inverse relationship between T and mass is that the particles are represented by the thermal de Broglie wavelength as the temperature is lowered; the wave packets begin to overlap and quantum effects characterize the system in a matter wave. When the mean de Broglie thermal wavelength of the particles approaches unity, quantum effects characterize the system. The BEC phase transition results from the particle wavelength spanning the interparticle distance, roughly the same point when exchange effects become important. We illustrate pertinent comparisons of BEC systems below [14]:

| Systems | Excitons | Polaritons |
|---|---|---|
| Effective mass $m^*/m_e$ | $10^{-1}$ | $10^{-5}$ |
| Critical temperature $T_c$ | 1mK -1K | $1 \rightarrow 300K$ |
| Thermalization time / Lifetime | 10 ps / 1ns ~$10^{-2}$ | $(1-10ps)/(1-10ps) = 0.1-10$ |

**3c. Kibble-Zurek Model (KZM) Theory [15].** Crucial to the discussion is an estimate as to how many random cosmic strings should there be as the universe rapidly cools in a rapid first order transition? Fortunately, condensed matter offers phase transitions where the relevant speed is not the speed of light, but for example, the second sound in a normal-to-superfluid second-order phase transition in helium. In this transition the order parameter (OP) can be used as a measure of the degree of progress since it ranges continuously from zero in the phase above the critical point to a finite number below it, and the order parameter susceptibility usually diverges. Landau exploited this continuous change property by placing two conditions on the free energy; the need to be analytic and obey the symmetry of the Hamiltonian. Close to a critical temperature, $T_c$, the free energy can be expanded in a Taylor series in the OP, here, the Ginsberg-Landau potential,

$$V(\psi) = \alpha(T)|\psi(\vec{r})|^2 + \tfrac{1}{2}\beta|\psi(\vec{r})|^4 \tag{4}$$

with $\psi$, the Bose-Einstein Condensate (BEC) complex order parameter, $\psi = \sqrt{\rho}e^{i\varphi}$, where $\rho$ is the fluid density and $\varphi$ is the phase. A complex order parameter permits formation of strings (or vortices). Near the critical temperature, $T_c$, the coefficient $\alpha = \alpha'(T-T_c)$ and both $\alpha'$ and $\beta$ are constants that can be determined by the specific heat data.

The coherence length, $\xi$, is the length of the order parameter, short for a "normal" state, and longer for a BEC (coherent) state. The coherence length diverges as the critical temperature is approached. The complex order parameter obeys the Gross-Pitaevskii equation (GPE):

$$i\hbar\frac{\partial \psi}{\partial t} = -\frac{\hbar^2}{2m}\nabla^2\psi + \alpha\psi + \beta|\psi|^2\psi \tag{5}$$

where m is the atomic mass. Applying eq. (5) to KZM and rescaling the characteristic equilibrium quantities, for example correlation length, $\xi_{eq} = \hbar/\sqrt{2m|\alpha|} = \xi_0/|\varepsilon|^{1/2}$, superfluid density, $|\Psi_{eq}|^2 = -\alpha/\beta$, and relaxation times, $\tau = \hbar/|\alpha| = \tau_0/\varepsilon$ where we assume $\varepsilon$ changes linearly since $\varepsilon = t/\tau_Q$, $\tau_Q$ being a quench time. The initial correlation length $\xi(t)$ of $\Psi$ has the same value as $\xi_{eq}(t)$ when approaching the phase transition from $t < 0$. The critical slowing forces $\xi_{eq}(t)$ to diverge, which suppresses the velocity of the spreading of the order parameter coherence. Consequently, the regions containing ordered phases begin to form independently in parts of the system. These disconnected domains continue growing and overlapping, eventually coalescing at time $\hat{t} = \sqrt{\tau_0 \tau_Q}$ as $\xi(t)$ approaches $\xi_{eq}(t)$. However, in different domains the order parameter phases $\theta$ are uncorrelated, leaving topological defects (such as vortices) at the *domain boundaries*. An estimated separation between defects can be estimated by the correlation length at $t = \hat{t}$, since

$$\hat{\xi} = \xi_{eq}(\hat{t}) = \xi_0 \left( \tau_Q / \tau_0 \right)^{1/4}, \tag{6}$$

resulting in an approximate defect density of $n_{def} \approx 1/\hat{\xi}^2$.

**3d. The Geodesic Rule and Defect Types [16].** Geodesics are the straightest curve in a manifold. Here, when bubbles of new phases are nucleated with random phases, these new regions follow a geodesic path that interpolates the new phase. A scenario for vortex formation in two space might be a simultaneous collision between three bubbles with phases of $2\pi/3$, 0, and $4\pi/3$. A geodesic rule would be to move from 0 to $2\pi/3$, then $4\pi/3$, so, we have completely covered the vacuum manifold once. Then, the value of the field is zero at the vortex core. Therefore, what happens if only two bubbles collide (a non-geodesic path); the bubbles can decay by nucleating vortices, or if a vortex is created transiently, a stable vortex could be formed. Alternatively, if the third bubble arrives before a vortex has been formed, the future of the vortex depends on the values of the bubble phases. When the sum of the three bubbles' phases generates a complete rotation around the vacuum manifold (to infinity), the vortex is **topologically stable**; otherwise, it is not.

**Defect Types.** In general, the matter symmetry and the type of phase transition determine the type of defect, so focusing on condensed matter, we will limit our discussion to three of the four basic types defined as:
Planar domain walls - A network of two-dimensional walls formed when a discrete symmetry is broken, resulting in partitions of the "universe" into cells.
Cosmic strings (**vortices**) - These strings result by the breaking an axial or cylindrical symmetry. A key feature is a set of non-simply connected minima surrounding the high energy state, resulting in a non-trivial winding number. **Our focus.**
Monopoles - Zero-dimensional objects which are diluted by inflation.
Textures – Delocalized topological defects that are unstable.

**4. Interacting Bose Condensations. a. Two Bose condensations.** A Josephson Junction (JJ) between a Bose gas trapped in a double-well magnetic trap with a high enough barrier can be solved using the GPE for each well [17, 18], and the classically forbidden region. A different number of atoms, $N_1, N_2$ will result in two different chemical potentials $\mu_1, \mu_2$, localized in their respective traps. Following Dalfovo et al., we take a linear combination of the two solutions to provide a solution of the time-dependent Schrödinger equation:

$$\psi(x,t) = \psi_1(x)\exp\left(-i\frac{\mu_1 t}{\hbar}\right) + \psi_2(x)\exp\left(-i\frac{\mu_2 t}{\hbar}\right) \qquad (7)$$

and we can show that $\psi_1$ and $\psi_2$ overlap significantly in the forbidden region. The current density is calculated and it has the form of a typical Josephson form:

$$I = I_0 \sin\frac{(\mu_1 - \mu_2)t}{\hbar} \qquad (8)$$

The Josephson current will result in an oscillation of the number of atoms in the two traps,

$$\frac{d}{dt}N_1 = -\frac{d}{dt}N_2 = -I_0 \sin\frac{(\mu_1 - \mu_2)}{\hbar} \qquad (9)$$

which has been experimentally verified [19].

**4b. Three Homogeneous Interacting Bose Condensations.** It is not obvious whether three cyclically and weakly coupled superfluids can result in a supercurrent since the current strength depends on initial relative phases and the strength of the coupling with the two adjacent superfluids [20, 21, 22]. An ansatz for the three order parameters is

$$\Psi(\vec{r},t) = \sum_i \psi_i(t)\Phi_i(\vec{r}) \qquad (10)$$

where $\Phi_i(\vec{r})$ represents each "domain" with a unique phase. Substituting eq. (xiii) into the Gross- Pitaevskii equation eq. (2), we find a cyclical linear relationship between adjacent

$$i\hbar\frac{\partial \psi_i}{\partial t} = \left(E_i + U_i|\psi_i|^2\right)\psi_i - K_{i,i-1}\psi_i - K_{i,i+1}\psi_{i+1} \qquad (i=1,2,3) \qquad (11)$$

complexes. While the equations for both cases are similar, only three or more supercurrents can create a topological entity. While the equations for both cases are similar, a key difference here is that the three supercurrents or cyclical Josephson currents (CJC) can enclose a topological entity (vortex) depending on the coupling constants and their initial relative phases. The size of the coupling constant is crucial since it must be small to avoid exciting the plasma state. In the weak coupling limit and assuming only stationary solutions produce CJC,

$$\langle j_c j_c \rangle_n = \frac{1}{(2\pi)^{n-1}} \sum_{k=0}^{n-1} \frac{2}{n}\left|\sin\left(\frac{2\pi k}{n}\right)\right|$$

$\langle j_c j_c \rangle_n$ decreases as $n \to \infty$. The conclusion is that most of the CJC's result from a small value of n at a stationary phase of $\phi_{1,2} = \phi_{2,3} = 2\pi/3$, so the value of $\langle j_{KZ} \rangle_3 = 0.577$ seems exaggerated. However, this value was calculated based on the maximum possible number of vortices. As Nozieres has pointed out, a BEC cannot be formed without interactions [23], so, if the separate Bose interactions were larger, vortex formation is more likely to happen, as in the example.

**4c. Merging Multiple Bose Condensates to Form Vortices.** There have been several interesting superfluid vortex experiments. Scherer et al. merged three dilute gaseous Bose condensates at both fast and slow rates to form quantized vortices, demonstrating a connection between interference, merging and vortex generation [24]. Ryu et al. created a persistent superfluid quantized rotation in a toroidal trap, initiated by the transfer of one unit $\hbar$ of orbital angular momentum [25]. Stable flow required a multiply connected trap and a condensate fraction of at least 20%. When two units of angular momentum were added, the vortices split into two singly charged vortices. While the KZM regards evolution as completely reversible and only topological winding numbers "remember" the scaling, the post-transition phase ordering being smoothed might be regarded a quantum analogue [26]. Our final exam focuses on the Meissner state where a planar, circular superconductor contains a hole with circumference, C [27]. The order parameter is a complex field $\psi = \rho e^{i\phi}$, with $|\rho|^2$, the density of Cooper pairs and phase $\phi$. Quenching the system from normal to superconducting prevents a uniform phase, so we define a topological winding number density, $n(x)2\pi = d\phi(x)/dx$, so the normalized magnetic flux through the hole is $n = \int_0^C n(x)dx = \Delta\phi/2\pi$. Eq. (6) follows, and a further assumption was $C \gg 2\pi\bar{\xi}$ and the probability of a single fluxoid trapping extrapolated to $f_1 \approx \langle n^2 \rangle \approx \frac{r}{\xi_0}\left(\frac{\tau_Q}{\tau_0}\right)^{-\sigma}$. The data suggested the exponent should be $-2\sigma$, suggesting small defects may be difficult to properly analyze in simulations.

**4c. Three Inhomogeneous Interacting Bose Condensations (CDW).** Recently Miller et. al. [28] have expanded earlier work, particularly by Maki [29] and Bardeen [30] on "depinning" a CDW trapped by impurities some twenty-five years ago. Bardeen showed experimentally that CDW's can tunnel through a potential barrier, like Josephson's discovery of cooperative quantum tunneling in superconductors [31]. Coleman broadened the description to include macroscopic quantum tunneling and decay of the "false vacuum" [32] that characterize instability in a scalar field relative to a lower energy state. We now know that two BEC's in proximity can use a Josephson Junction (JJ) to exchange particles.

For the moment, we will assume that this theory allows the possibility of a B850 complex CDW in an excited state (metastable well) to tunnel to adjacent unexcited B850 rings or a ground state B850 to tunnel a lower potential well (B875). The event nucleates a bubble of a "true vacuum" contained within a soliton description. The quantum charged $(\pm 2e)$ solitons, delocalized in the transverse directions, use Josephson tunneling to move to another CDW chain. Following Miller, we consider nucleated droplets of kinks and anti-kinks with a $\pm 2e$ charge as quantum fluids with quantum delocalization between CDW chains. The model proposed relates a vacuum angle $\theta$ to displacement charge Q between inter-chain contacts $\theta = 2\pi(Q/Q_0)$, where the potential energy of the ith-chain is,
$$u_i(\varphi_i) = 2u_0\left[1 - \cos\varphi_i(x)\right] + u_E\left[\theta - \varphi_i(x)\right]^2$$
We ignore the first term, the periodic DW pinning energy, to focus on the quadratic electrostatic energy from the net charge displacement; graphs of u vs. $\theta$ for $\varphi$ (approximately $2\pi n$) when the energy is minimized (Fig 4). Tunneling is coherent into the next well (chain) via tunneling matrix element T as each parabolic branch crosses the next instability point, $\theta = 2\pi(n + 1/2)$.

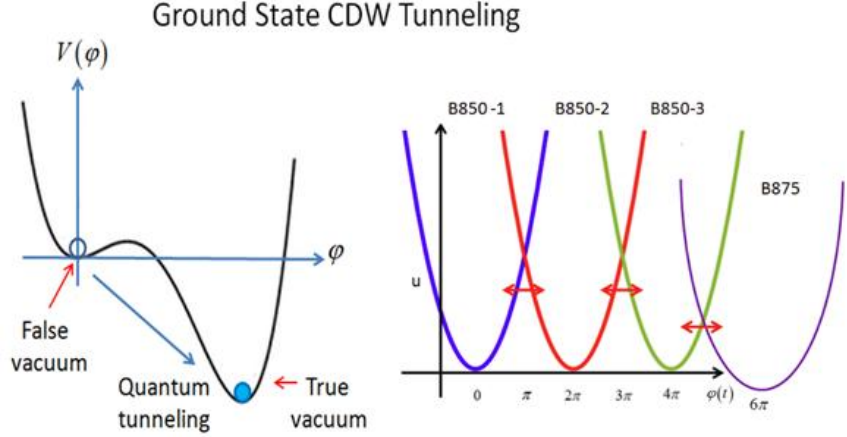

Figure 4. (left) Both tunneling to a lower energy potential well and (right) to a series of identical B850 complex rings can take place. Eight B850 complexes surround the B875-RC complex. If the population changes, say, in B850-2, CDW tunneling could rapidly re-equilibrate the population.

Since there are eight C850 CDW rings surrounding B875, also a CDW (Section 1), multiple vortices are very possible since the interaction term $K_{CDW}$ can be considerably larger than the plasma energy level restriction in the superfluid case (section 4b). Based on the eight possible vortices in the B850 / B875 complex, it seems likely that at least two vortex tubes are present. In fact, the value of $K_{CDW}$ could be several hundred meV's, since the CDW is stable at room temperature, so there could be as many as eight (or nine including B875) coherent rings and vortex tubes.

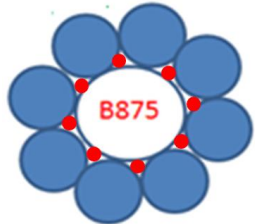

Figure 5. Eight circular B850 CDW's surround the B875 CDW. The red dots indicate the location of possible vortex tubes.

**4d. More than Three CDW's.** A model system in this regard could be the quasi-1D Tritellurides, $TbTe_3$, where trARPES (time and angle resolved photoemission spectroscopy) experiments after ultrafast optical excitation revealed a coherent oscillation of the occupied CDW band [6, 7, 33, 34], and the Peierls distortion energy gap $\Delta(\vec{k})$ was measured by examining the occupied ground state and unoccupied excited state. Here, trARPES (time and angle resolved photoemission spectroscopy) experiments after ultrafast optical excitation revealed a coherent oscillation of the occupied CDW band at about 2.3 THz. The oscillation modulated the CDW's peak position, amplitude, and spectral width. The unoccupied band also had a downshift and the two band shifts reduced the gap size. The Peierls' distortion energy gap $\Delta(\vec{k})$ was measured by examining the occupied ground state and unoccupied excited state.

Since the complex order parameter can be expressed as $\Delta = |\Delta|e^{i\varphi}$, the variation of the magnitude of the gap (called an "amplitudon") was measured, while the phase (a "phason") remained constant [7]. Another experimental scheme measured both the gap and the phases of both bands using (trARPES) in conjugation with a three-pulse probe with various appropriate time delays that can repopulate the upper, normally unoccupied CDW band over extended times completed the dynamic measurements of the CDW gap in tritellurides.

The experimental set-up is a femto-to- picosecond three-pulse excitation scheme. The first pulse excited and created the coherent amplitude mode oscillations; the second mode repopulated the unoccupied states and the third probe pulse, the bands. A Lorentzian line fit of the peaks of the upper and lower bands show oscillations of the same magnitude in both bands that are clearly anti-correlated. Both single-particle and collective modes can be tracked with femtosecond resolution through various phase transition, hence rapid topological defects can be followed in an ultrafast experiment that opens a new opportunity to explore the Kibble-Zurek mechanism.

## 5. a. Polariton States and Their Possible Use in the PLHC. [35, 36]

A polariton is a two-level state created by the superposition of a photon and another boson with a dipole such as an exciton. Polaritons were discovered years ago, but they were difficult to study. Microcavities could overcome this barrier, but it took years to fabricate a microcavity with the dimensions lower than $\lambda/2$. The polariton interaction Hamiltonian is the dot product of the dipole energy (exciton) in the polarization electric field of a photon, $\vec{E}$,

$$H_{int} = -\int d^3 r \vec{P} \cdot \vec{E}$$

This technology enabled a much larger polariton density that was further increased by the recognition that the large oscillator strength of organic materials leads to a much larger Rabi separation between the levels (1.2 eV). Both advances allow a macroscopic number of polaritons to occupy the same state so a BEC forms even though these quasiparticles are *inherently in non-equilibrium states*. Nonetheless, measured polariton properties include: coherence, off-diagonal long-range order (ODLRO), superfluidity, Bogoliubov excitations, quantized vortices, and solitons (topological "lumps of energy") that could tunnel from excited B850 complexes or CDW's [14].

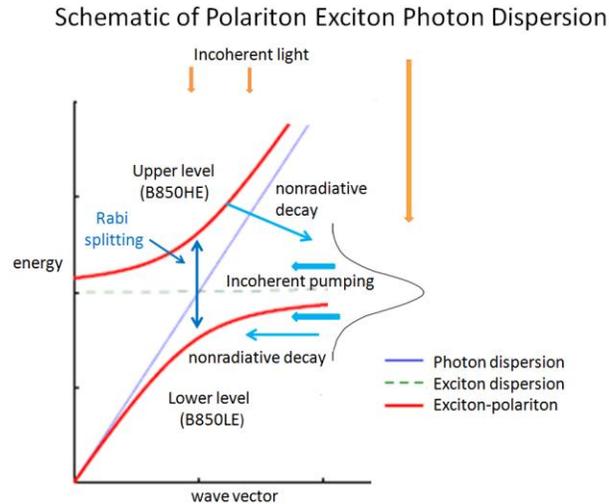

Figure 6. An electric field pulls an exciton comprised of an electron and a hole in different directions affecting the dielectric constant of the medium. The resonance frequency of the exciton is not constant, resulting in variable branches of the exciton-polariton, the strength of which is measured by the Rabi splitting between the two branches. The exciton oscillator strength determines the Rabi splitting. *The eigen-modes mix only in the region of the crossover.*

**b. Why Polaritons in the PLHC?** An exciton-polariton should be present simply because it seems to be the only suitable coherent structure at room temperature to absorb incident photons completely (99%) and deliver fast, near perfect, energy transfer over significant distances. The B800-B850-B875 supercomplex ($B^3SC$) should be able to store energy using the Jaynes-Cummings effect. An examination of the dipole structure of the ($B^3SC$) shows attributes of a nanopillar, normally a synthetic device that captures over 99% of incident light [4, 5] and coherently transfers the energy to Reaction Centers, either near or far using (coherent) solitons. Even assuming a nanopillar, the detailed physical description of a PHLC contains several anomalies, which we can rationalize with our model, such as:

1. Dostal et al using 2D spectroscopy observed a "rapid loss of excitonic coherence between structural subunits" [37]. After the rapid transfer, the chemical potential becomes zero, hence transfer is in equilibrium.
2. The ability of *uphill* energy transfer from B875 to B850 [38, 39, 40, section 3b] could result from a highly excited B875 matter wave peak becoming transiently overlapping a non-excited B850 complex. More likely though is a mixing of the B875 and B850 eigenmodes in the crossover region.
3. The non-linear splitting of the B850 band by high and low light intensities (HL and LL) into high (B850HE) and low energy (B850LE) bands to describe the spectra properly [39, 40]. This non-linear splitting could be a Rabi splitting of a polariton.
4. A polariton has the ability to extend its length to the order of the size of the cavity mode [41], and to generate solitons [42]. Josephson Junctions are also present, so it seems that coherent entities can transfer energy through coherent "links" [43].
5. When related bacteria expand the B850 ring, it is two units of B850 that are added to the ring, presumably to maintain the Peierls' distortion; a single B800 is also added between two B850 units.
6. The B850 UL/LL could support energy storage via Jaynes-Cummings effect [44].
7. Coles et al [45] have formed a polariton from strong coupling between a confined optical cavity mode and a photosynthetic bacteria chlorosomes, so polariton formation is possible
8. Entanglement seems possible in the ground and excited states (fig 8).

**6. Possible Experiments. a. Ground State.** The three-pulse system (D, destruction, P-p, Pump-probe) [6, 7] could focus on a single ground state CDW, two adjacent CDW's, two non-adjacent CDW's to monitor and/or suppress the susceptibility. A second probe might be useful. It there a real possibility of eight vortex tubes being present? In a magnetic field will there be current rotation of the eight B850 CDW's or will they flow in opposite rotations so the overall effect of the eight B850 complex be "neutral". However, if two adjacent B850's have a current, can it rotating in opposite directions and possibly create a flux loop perpendicular to the plane of the rings?

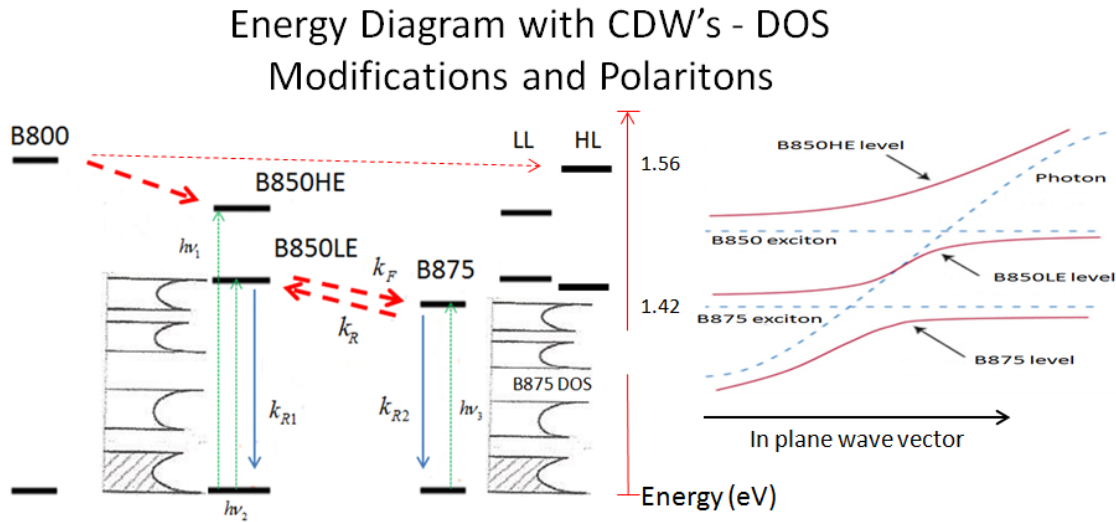

Figure 7. Excited state energy levels and band gaps for B800 molecules (left), then eight B850's surrounding the B875/RC along with possible polariton structures for these states. Most bosons are restricted to the region of $\Delta p = 0$; polaritons can have considerable freedom in this regard, covering distances to $50 \mu m$ or more [40]. The breaks in the 1D CDW's is due to a band gap and the Peierls' distortion gap and its symmetry partner. The LL and HL illustrate the non-linear splitting of the B850 band (center top). In LL there would probably be only two polariton levels and B800 could contribute incoherent energy; in HL B800 could convert into transfer of coherent energy to B850HL.

**b. Excited States.** Many of the ground state experiments can be expanded and repeated in the polariton / CDW complex at low light and high light, these "normal" light being a fourth energy source. There is considerable expertise in microcavity spectroscopy and nanocavity work is progressing, all of which is specifically designed to avoid or selectively include the continuum.

**7. Summary and Conclusions.** Have we answered the question in the article title? If the array of rings surrounding the B875 / RC are CDW's based on their "dimer" structure, then the answer is yes. Each circular CDW requires a separate symmetry breaking (SB) [1]. Are these SB's correlated? Probably not when the structure was assembled, but almost certainly later. The time scale of PLHC events ranges from pico to microseconds, so a key question is on a high light day, how is the energy distributed so that the various sensitive entities avoid photo bleaching? Based on the discussion in section 5b, we suggest that the coherent entities (chemical potential $\mu = 0$) and using what we believe is the dominate connecting transfer mechanism, solitons, rapidly smooth the chemical potential over the complex [37]. Kim et al. have recently observed a soliton transfer between CDW's following the Bardeen- Miller mechanism [46], and Sich et al. report that giant optical non-linearity enables the formation of mesoscopic polariton solitons numbering in the hundreds [47].

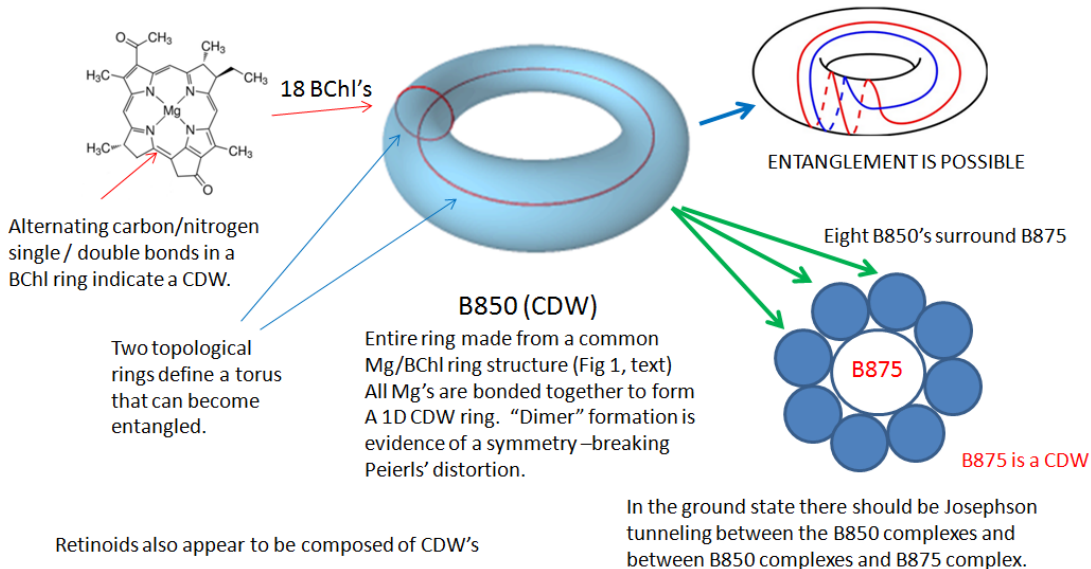

Figure 8. Circular CDW's in a magnetic field have been shown to quantize the magnetic flux in units of $\Phi = nh/2e = n\Phi_0$ where n is the number of flux. The Meissner effect is a rotation of charge in a SC or CDW that opposes the external flux, permitting only the flux in the fundamental units of $\Phi_0$ (or $h/2e$). In principle BChl could generate an n, a B850 complex, n', and the eight B850 surrounding B875, n''.

**Acknowledgement.** RHS is grateful to NHM for changing his focus some twenty-five years ago.